\documentclass[
floatfix, 
pre,
aps, 
preprint, 
]{revtex4-1}

\pdfoutput=1

\usepackage{graphicx}
\usepackage{amssymb}
\usepackage{amsmath}
\usepackage[utf8]{inputenc}
\usepackage[T1]{fontenc}
\usepackage{cancel}
\usepackage[usenames,dvipsnames]{color}
\usepackage{psfrag}
\usepackage{epsfig}
\usepackage{bbm}
\usepackage{bm}
\usepackage{dsfont}
\usepackage{soul}
\usepackage{colonequals}
\usepackage{float}
\usepackage{enumitem}

\usepackage{tikz}
\usepackage{pgfplots}
\usetikzlibrary{pgfplots.groupplots}
\usetikzlibrary{shapes}

\usepackage[colorlinks, allcolors = blue] {hyperref}
\makeatletter
\renewcommand*{\eqref}[1]{%
  \hyperref[{#1}]{\textup{\tagform@{\ref*{#1}}}}%
}
\makeatother

\usepackage[caption=false]{subfig}

\newcommand{\bs}[1]{\boldsymbol{#1}} 
\newcommand{\makeEffectiveAverageDensitySymbol}{\bar{\rho}_{\text{eff}}} 
\newcommand{\makeCriticalEffectiveAverageDensitySymbol}{\bar{\rho}_{\text{eff,c}}} 

\begin{document}
\title{Freezing and re-entrant melting of hard discs in a one-dimensional potential:\\
Predictions based on a pressure-balance equation
} 

\author{Alexander Kraft}
\email{alexander.kraft@tu-berlin.de}
\author{Sabine H.~L.~Klapp}
\email{sabine.klapp@tu-berlin.de}

\affiliation{
  Technische Universit\"at Berlin,
  Institut f\"ur Theoretische Physik,
  Straße des 17. Juni 135, 
  10623 Berlin,
  Germany}
\
\date{\today}

\begin{abstract}
We investigate theoretically the freezing behaviour of a two-dimensional (2D) system of hard discs on a one-dimensional (1D) external potential (typically called laser-induced freezing).
As shown by earlier theoretical and numerical studies, one observes freezing of the modulated liquid upon increase of the substrate potential amplitude, and re-entrant melting back into the modulated liquid when the substrate potential amplitude is increased even further.
The purpose of our present work is to calculate the freezing and re-entrant melting phase diagram based on information from the bulk system.
To this end, we employ an integrated pressure-balance equation derived from density functional theory [Phys. Rev. E \textbf{101}, 012609 (2020)].
Furthermore, we define a measure to quantify the influence of registration effects that qualitatively explain re-entrant melting.
Despite severe approximations, the calculated phase diagram shows good agreement with the known phase diagram obtained by Monte Carlo simulations.
\end{abstract}

\maketitle

\section{Introduction \label{SEC:INTRO}}

Hard body interactions represent the simplest form of interaction between particles and are frequently used as reference systems in the statistical-mechanical description of classical many-body systems~\cite{HansenMcDonald}.
The earliest work on the freezing of hard bodies dates back to the seminal computer simulation study of Alder and Wainwright~\cite{Alder1957} for hard spheres in three dimensions (3D). They established the concept of an (entropy-driven) freezing transition of particles that purely interact via repulsion~\cite{HansenMcDonald}.
Whereas the phase diagrams of homogeneous (bulk) systems of hard spheres and hard discs are well understood in 3D~\cite{Hoover1968, Pusey1986, HansenMcDonald} and 2D~\cite{Bernard2011, Engel2013, Thorneywork2017}, the theoretical prediction of the hard-body phase behaviour in complex geometries or inhomogeneous external potentials, remains difficult.

In this work, we are interested in hard spheres confined to 2D (hard discs) and subjected to a 1D periodic substrate potential, here taken as a sine substrate.
The phenomenon of freezing of a 2D colloidal suspension on a 1D periodic substrate is commonly denoted as laser-induced freezing (LIF) and was first  discovered experimentally by Chowdhury, Ackerson, and Clark~\cite{Chowdhury1985} in a 2D monolayer of charged spherical particles subjected to a commensurate 1D periodic light field.
This observation led to a series of studies by theory~\cite{Chakrabarti1994, Das1998, Das1999a, Frey1999, Radzihovsky2001, Rasmussen2002, Chaudhuri2004, Nielaba2004, Chaudhuri2006, Luo2009}, computer simulations~\cite{Loudiyi1992b, Chakrabarti1995, Das1999a, Das1999b, Das2001, Strepp2001, Strepp2002, Strepp2003, Chaudhuri2004,Chaudhuri2005, Chaudhuri2006, Buerzle2007, Luo2009} and experiments~\cite{Loudiyi1992a, Wei1998, Bechinger2000, Bechinger2001, Baumgartl2004}. 
From the theoretical side, a major step towards an understanding of the full LIF scenario was provided by the work of Frey, Nelson, and Radzihovsky~\cite{Frey1999, Radzihovsky2001}.
They extended the concept of dislocation-mediated melting in 2D described by KTHNY theory~\cite{Kosterlitz1973, Halperin1978, Nelson1979, Young1979} to the presence  of 1D periodic substrates. 
Extensive Monte Carlo (MC) simulation studies~\cite{Strepp2001, Strepp2002, Strepp2003, Buerzle2007} later confirmed their results.

However, whereas the physical concepts underlying LIF are understood for more than two decades, it remains difficult to make {\em quantitative} theoretical predictions for LIF in different model systems (i.e., different interaction potentials). 
The LIF phase diagram of hard discs has been obtained through extensive MC simulation studies~\cite{Strepp2001} and was studied theoretically~\cite{Chaudhuri2004, Chaudhuri2006} based on renormalization group flow equations, with input from constrained MC simulations.
The simplest (and to our knowledge the only) \emph{purely} theoretical prediction for the phase diagram is based on density functional theory~(DFT)~\cite{Rasmussen2002}. However, the resulting diagram differs qualitatively (and, thus, also quantitatively) from the one obtained from MC simulations~\cite{Strepp2001} due to severe approximations for the excess free energy.
Here we propose another strategy.

In previous work, we developed a framework based on a pressure-balance equation~\cite{Kraft2020a} to theoretically predict the LIF of ultrasoft particles on two different substrate types (cosine and Gaussian). The results agreed well with numerical calculations based on DFT~\cite{Evans1979,Evans1992}.
The core idea of our approach is that the modulation by the 1D periodic substrate leads to an increase of a (suitably defined) effective average density close to the potential minima.
This region can be characterized by a width $L_c$ which is smaller than the substrate periodicity $L_s$.
The developed framework~\cite{Kraft2020a} allows to calculate $L_c$ as function of the system parameters (such as the average system density $\bar{\rho}$) and as function of the substrate parameters (such as the potential strength $V_0$).
Typically, $L_c$ decreases with increasing $V_0$ at fixed overall density.
The resulting increase of effective average density $\makeEffectiveAverageDensitySymbol$ within this region of confinement then leads to LIF.
One goal of our present work is to utilize this strategy~\cite{Kraft2020a} to predict LIF in a hard-disc system.

Besides freezing, one observes for various types of systems~\cite{Nielaba2004} exposed to 1D periodic substrates a re-entrant melting. Here, the liquid first freezes at some potential strength $V_0$ and then melts again for sufficiently large values of $V_0$.
The re-entrant melting was attributed by Wei \textit{et al.}~\cite{Wei1998} to a reduced "registration" of particles in neighbouring potential minima, caused by a decrease of fluctuations perpendicular to the standing-wave pattern.
This provides an intuitive understanding of re-entrant melting phenomenon. 
However, it is unclear to which extent the registration has to be reduced to induce re-entrant melting.
In the present work, we therefore define a measure to quantify the registration effect introduced in Ref.~\onlinecite{Wei1998}.
Extending our framework by this registration measure (which is based on $L_c$), we can make a prediction for re-entrant melting.
Altogether, our work provides a recipe how to calculate the 2D phase diagram of hard discs on a 1D periodic substrate \emph{based on information from the bulk system and known limiting behaviours}.

This article is organized as follows:
In Sec.~\ref{Sec:Calculation_of_the_phase_diagram}, we introduce our model and summarize key steps of our theoretical prediction~\cite{Kraft2020a}. By this, we make predictions for the onset of LIF and re-entrant melting.
In Sec.~\ref{Sec:Discussion_of_the_phase_diagram}, we discuss our calculated phase diagram and compare it with the phase diagram obtained by MC simulations~\cite{Strepp2001}. We summarize our findings and outline directions for future research in Sec.~\ref{Sec:Conclusion}.

\section{Calculation of the phase diagram \label{Sec:Calculation_of_the_phase_diagram}}

In this work, we calculate the phase diagram for hard discs in 2D (located along the $x$-$y$ plane) with diameter $\sigma$ and interaction potential $V(r)$,
\begin{align}
\label{Eq_hard_disc_interaction_potential}
V(r) =	\begin{cases}
					\infty & r\leq\sigma\\
					0 & r>\sigma,\\
		\end{cases}	
\end{align}
on an external sine potential along the $x$-direction, that is,
\begin{align}
V_{\text{ext}}(x) = V_0 \sin\left(\frac{2 \pi x}{L_s}\right),
\label{Eq_sine_substrate_potential}
\end{align}
with periodicity $L_s$ and potential amplitude $V_0$ (thus with potential difference $2 V_0$).
The substrate periodicity $L_s$ is given in units of the nearest neighbour distance $a$ of the solid, i.e. $a = \left(2/\sqrt{3}\bar{\rho}\right)^{\frac{1}{2}}$ with average system density $\bar{\rho} = N/A$, number of particles $N$, and system area $A$. Specifically, we set $L_s / a = \sqrt{3}/{2}$. This choice of the substrate potential and periodicity is in agreement with Ref.~\cite{Strepp2001}, where the phase diagram was calculated by MC simulations.
In particular, the choice for $L_s$ ensures that the 1D periodic substrate is commensurate~\cite{Bechinger2007} with the hexagonal solid.

\begin{figure}
	\centering
	
	\includegraphics[width=0.4\textwidth]{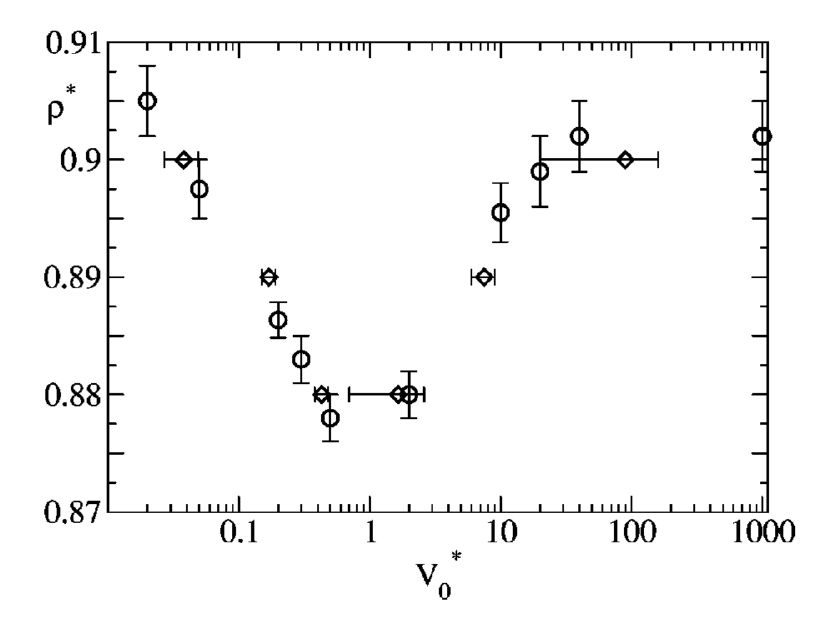}

	\caption{The phase diagram as obtained in Monte Carlo simulations in Ref.~\cite{Strepp2001}, where $V_0^* = \beta V_0$ and $\rho^* = \bar{\rho} \, \sigma^2$ in our nomenclature. Symbols denotes the phase boundary between the locked floating solid phase (above) and the modulated liquid (below the symbols). From Strepp \textit{et al.}, 2001.
	}
	\label{Fig_phase_diagram_Strepp2001}
\end{figure}

The phase diagram obtained in Ref.~\cite{Strepp2001} is shown in Fig.~\ref{Fig_phase_diagram_Strepp2001}.
Below the phase boundary, the system displays a so-called modulated liquid phase. 
Here, the density profile is modulated by the external potential~[see Eq.~\eqref{Eq_sine_substrate_potential}] along the $x$-direction, but is constant along the $y$-direction. Thus, $\rho(x,y) = \rho(x)$. For the so-called locked floating solid which appears above the phase boundary in Fig.~\ref{Fig_phase_diagram_Strepp2001}, the density profile  $\rho(x,y)$ is truely two-dimensional; it reflects the formation of a hexagonal solid that is commensurate with the substrate.
As indicated by the phase boundary in Fig.~\ref{Fig_phase_diagram_Strepp2001}, freezing on a 1D periodic substrate shows two prominent features.
First, upon increasing $V_0$ from zero at fixed density $\rho^*= \bar{\rho} \sigma^2$, one observes freezing below the bulk freezing density ($\bar{\rho}_f \sigma^2 = 0.93$~\cite{Thorneywork2017}).
Second, there is a range of densities $\rho^*$ where laser-induced freezing is followed by re-entrant melting upon further increase of the potential amplitude $V_0$.
The goal of our work is to reproduce these two phenomena based on a theoretical framework~\cite{Kraft2020a}, which involves information from the bulk system.


\subsection{Details of the theoretical framework\label{Sec:Details_of_the_theoretical_framework}}
In our previous work on LIF of ultrasoft particles~\cite{Kraft2020a}, which involved numerical DFT calculations, we found evidence that LIF can be seen as a density-driven transition induced by the increase of the (suitably defined) effective average density, $\bar{\rho}_{\text{eff}}$, in the vicinity of the potential minima. 
We then assumed that there is a "critical" threshold $\bar{\rho}_{\text{eff,c}}$ which, upon exceeding, leads to spontaneous symmetry breaking, that is, a change of the density profile from $\rho(x)$ to $\rho(x,y)$.

The actual calculations are based on an \emph{ansatz} for the density profile around the potential minimum, say $x=0$. (For notational convenience, we assume a symmetric and appropriately shifted external potential such that $V_{\text{ext}}(x) = V_{\text{ext}}(-x)$, $V_{\text{ext}}(0) = 0$). The ansatz has the form of a rectangular function with width $L_c$ and height $\bar{\rho}_{\text{eff}}$, namely,
\begin{subequations}

	\begin{align}
		\rho(x) 	&= \bar{\rho}_{\text{eff}} \, \text{rect}\left(\frac{x}{L_c}\right) = \begin{cases}
        	                                                                 \bar{\rho}_{\text{eff}} & 								\text{if }  |x| \leq \frac{L_c}{2} \\
        	                                                                 0			&	\text{else}.
        	                                                                \end{cases} 
        	                                                                \label{Eq_effective_density_rectangular_ansatz}							\\
\intertext{where}
		\bar{\rho}_{\text{eff}} &= \bar{\rho} \, \frac{L_s}{L_c}.
		\label{Eq_redistribution_of_particles}
	\end{align}
	\label{Eq_effective_density_ansatz}
\end{subequations}
The parameters $L_c$ and $\bar{\rho}_{\text{eff}}$ are related by the conservation of particles [see Eq.~\eqref{Eq_redistribution_of_particles}].
The idea behind the ansatz~\eqref{Eq_effective_density_ansatz} is that the majority of particles are effectively located within a "confining" region (around the substrate minima) which is smaller than the substrate periodicity itself. 
We then developed a framework to calculate such a "confining length" $L_c$ and consequently the effective average density $\bar{\rho}_{\text{eff}}$ [see Eq.~\eqref{Eq_effective_density_ansatz}] as function of the system parameters, such as the average system density $\bar{\rho}$, and of the external potential, particularly the potential amplitude $V_0$. 
To do so, we started from an integrated version of the (exact) stress balance equation and then performed approximations involving the rectangular density profile~\eqref{Eq_effective_density_rectangular_ansatz} and a corresponding rectangular pressure profile (for a summary, see Appendix \ref{Sec:Background_of_our_framework}). 
This leads us to the equation
\begin{align}
		\label{Eq_averaged_stress_within_effective_density_ansatz_variant3}
	2 \, p(\bar{\rho}_{\text{eff}})    + \tilde{I}_{\bs{\tau}}( \bar{\rho}_{\text{eff}}, L_c  )
	&= 2\, \bar{\rho}_{\text{eff}}  V_{\text{ext}}\left(\frac{L_c}{2} \right)
\end{align}
for an effective bulk liquid with density $\makeEffectiveAverageDensitySymbol$. 
In Eq.~\eqref{Eq_averaged_stress_within_effective_density_ansatz_variant3}, $p(\bar{\rho}_{\text{eff}})$ is the bulk pressure at density~$\makeEffectiveAverageDensitySymbol$, and $\tilde{I}_{\bs{\tau}}( \bar{\rho}_{\text{eff}}, L_c  )$ is a correction term due to inhomogeneity, both of which arise when decomposing~\cite{Long1961MechanicsOfSolidsAndFluids} the stress tensor $\bs{\sigma}$ according to $\bs{\sigma} = -p\, \bs{1} + \bs{\tau}$. 
We employed a prescribed threshold value for $\bar{\rho}_{\text{eff}}$ taken from the bulk system.
We remark that the derivation of Eq.~\eqref{Eq_averaged_stress_within_effective_density_ansatz_variant3} does not require an explicit choice of the particle interaction or correlation functions.
Both are encapsulated within $p$ and $\tilde{I}_{\bs{\tau}}$, which allows to transfer the previously developed LIF prediction to other systems, as we will demonstrate in this work.

Our starting point will be again Eq.~\eqref{Eq_averaged_stress_within_effective_density_ansatz_variant3}, which we rewrite in the form
\begin{align}
\label{Eq_abbreviated_EOS_form_averaged_stress_within_effective_density_ansatz}
	 Z(\bar{\rho}_{\text{eff}}) + \tilde{\Gamma}(\bar{\rho}_{\text{eff}}, L_c) = \beta V_{\text{ext}}\left(\frac{L_c}{2} \right),
\end{align}
where we identified the compressibility factor $Z=  {\beta \,  p(\bar{\rho}_{\text{eff}})  }/{\bar{\rho}_{\text{eff}}}$, and we defined 
\begin{align}
	\tilde{\Gamma} \equiv { \beta \, \tilde{I}_{\bs{\tau}}( \bar{\rho}_{\text{eff}}, L_c  )}/{(2 \bar{\rho}_{\text{eff}})}.
	\label{Eq_Define_Gamma_Tilde}
\end{align}
We note that $\makeEffectiveAverageDensitySymbol$ depends on the system density $\bar{\rho}$ and $L_c$ through Eq.~\eqref{Eq_redistribution_of_particles}; therefore $\tilde{\Gamma} = \tilde{\Gamma}(\makeEffectiveAverageDensitySymbol(\bar{\rho}, L_c), L_c)$.
Further, the compressibility factor~$Z$ corresponds to a bulk system of density~$\makeEffectiveAverageDensitySymbol$. 
The latter is usually known for the system of interest, as the (homogeneous) bulk system is typically studied \emph{before} proceeding to inhomogeneous systems. 
The quantity $\tilde{\Gamma}$ [see Eq.~\eqref{Eq_Define_Gamma_Tilde}] is generally unknown. Here we will make approximations that allow us to calculate the LIF phase diagram \emph{solely} from bulk quantities.


\subsection{Prediction for the onset of LIF \label{Sec:Prediction_for_the_onset_of_LIF}}

We now propose a strategy how to use existing numerical or experimental data as an input for the quantities appearing in Eq.~\eqref{Eq_abbreviated_EOS_form_averaged_stress_within_effective_density_ansatz}. We specialize on a hard disc system. The equation of state for hard discs was determined experimentally in Ref.~\onlinecite{Thorneywork2017}, yielding the compressibility factor 
\begin{align}
 Z = \frac{\beta \,  p  }{\bar{\rho}} = 
				\begin{cases}
				{1}/{(1-\phi)^2}							 &, 0 \leq \phi \leq \phi_{\text{lc}}  \\
				  1/(1 -\phi_{\text{lc}} )^2 = \text{const}					 &, \phi_{\text{lc}} \leq \phi \leq \phi_{\text{hc}} \\
				{a}/{(\phi_{\text{cp}} - \phi)}						&,  \phi_{\text{hc}} \leq \phi \leq \phi_{\text{cp}}
				\end{cases}
				\label{Eq_Compressibility_factor_bulk_system_hard_disc}
\end{align}
where $\phi = \bar{\rho} \sigma^2 \, \pi/4$ denotes the packing fraction, $\phi_{\text{lc}} = 0.68$ and $\phi_{\text{hc}} = 0.70$  are the liquid and hexatic phase coexistence packing fractions, respectively, 
$\phi_{\text{cp}} = \pi /\sqrt{12}$ is the hard disc close packing fraction and $a= {(\phi_{\text{cp}} - \phi_{\text{hc}})   }/{(1-\phi_{\text{lc}})^2}$ is simply a number.
In Eq.~\eqref{Eq_Compressibility_factor_bulk_system_hard_disc}, the expression for the range $\phi \leq \phi_{\text{lc}}$ stems from scaled particle theory for the liquid phase~\cite{Thorneywork2014, Helfand1961}. The high packing fraction branch $\phi \geq \phi_{\text{hc}}$ is a semi-empirical fit~\cite{Salsburg1962} of experimental data of Ref.~\cite{Thorneywork2017}.
We note that the hard disc solid melts (or in reverse, freezes) via unbinding of dislocation pairs 
at the hexatic-solid transition at \mbox{$\phi_{f} =  0.73$}
according to KTHNY theory~\cite{Kosterlitz1973, Halperin1978, Nelson1979, Young1979}, without any signatures in the equation of state.
The corresponding reduced density \mbox{$\rho^*_f = \bar{\rho}_f \sigma^2 = \phi_f \,4/\pi$} then follows as $\rho^*_f = 0.93$.

The remaining task is to construct an approximation for $\tilde{\Gamma}$ [see Eq.~\eqref{Eq_Define_Gamma_Tilde}] for the hard disc system. 
To this end we consider two limiting cases: (i) the limit of densities close to the bulk freezing transition (i.e. $\bar{\rho} \to \bar{\rho}_f$) and (ii) the limit of vanishing densities ($\bar{\rho} \to 0$).

(i) We consider a confined system with an average density $\bar{\rho}$ somewhat below the bulk freezing density, $\bar{\rho}_f$.
In the limit $\bar{\rho} \to \bar{\rho}_f$, the potential amplitude $V_0$ required to induce LIF goes to zero.
This is known from the MC phase diagram~\cite{Strepp2001} and it is also consistent with our expectation: At bulk density, the system does not "need" a substrate to freeze.
Turning now to Eq.~\eqref{Eq_abbreviated_EOS_form_averaged_stress_within_effective_density_ansatz}, we see that, for vanishing external potential, the right side vanishes.
This implies that the left side of Eq.~\eqref{Eq_abbreviated_EOS_form_averaged_stress_within_effective_density_ansatz} must vanish as well, yielding
\begin{align}  
	\lim_{\bar{\rho} \to \bar{\rho}_f}   \tilde{\Gamma}(\makeEffectiveAverageDensitySymbol(\bar{\rho}, L_c), L_c) = - Z(\bar{\rho}_f),
	\label{Eq_limit_close_to_bulk_phase_boundary}
\end{align}
where we explicitly highlighted the dependency of $\makeEffectiveAverageDensitySymbol$ on the system density $\bar{\rho}$ [see Eq.~\eqref{Eq_redistribution_of_particles}].

(ii)~We can extract a further limiting case for $\tilde{\Gamma}$ in the limit of vanishing density, i.e., $\bar{\rho} \to 0$.
Physically, we simply expect that since there are no particles, the correction term in the stress tensor due to inhomogeneity vanishes, and thus
\begin{equation}
\lim_{\bar{\rho}\to 0}  \tilde{\Gamma}(\bar{\rho}, L_c) = 0,
	\label{Eq_limit_of_vanishing_density}
\end{equation}
with $\tilde{\Gamma}(\bar{\rho}, L_c)$ being the compact notation for the dependency $\tilde{\Gamma}(\makeEffectiveAverageDensitySymbol(\bar{\rho}, L_c), L_c)$.
We note that Eq.~\eqref{Eq_limit_of_vanishing_density} is consistent with our starting point, Eq.~\eqref{Eq_averaged_stress_within_effective_density_ansatz_variant3}. For $\bar{\rho} \to 0$, $\makeEffectiveAverageDensitySymbol$ vanishes as well, and so does $p(\makeEffectiveAverageDensitySymbol)$. Combing this with the zero at the right side of Eq.~\eqref{Eq_averaged_stress_within_effective_density_ansatz_variant3}, one arrives at Eq.~\eqref{Eq_limit_of_vanishing_density}.

There remains the question how $\tilde{\Gamma}$ depends on $\bar{\rho}$ in between the limits considered in Eqs.~\eqref{Eq_limit_close_to_bulk_phase_boundary} and \eqref{Eq_limit_of_vanishing_density}.
One possible approach is to just interpolate between these two limiting behaviours.
Here we use a simple ansatz for $\tilde{\Gamma}$ which satisfies both limits, namely
\begin{equation}
 \tilde{\Gamma} = - Z(\rho_f) \left(\frac{\bar{\rho}}{\bar{\rho}_f}\right)^n. 
 \label{Eq_Interpolation_for_GammaTilde}
\end{equation}
We stress that there is no a priori justification for the ansatz~\eqref{Eq_Interpolation_for_GammaTilde} for densities $0<\bar{\rho}<\bar{\rho}_f$.
However, the ansatz turns out to be surprisingly robust.
In particular, as shown in Appendix~\ref{Sec:Impact_of_technical_parameters}, the results are not very sensitive to $n$. For simplicity, we therefore set $n=1$.

Based on the expressions for $Z$ and $\tilde{\Gamma}$ we can now calculate the onset of LIF as shown in our previous work~\cite{Kraft2020a}. The idea is to prescribe a threshold value $\bar{\rho}_{\text{eff}} = \bar{\rho}_{\text{eff,c}}$ which the effective average density has to exceed at the LIF phase transition.
The corresponding confining length then follows from Eq.~\eqref{Eq_redistribution_of_particles} as
\begin{align}
 L_c &=  \frac{\bar{\rho}}{\bar{\rho}_{\text{eff}}}   L_s.
\end{align}
For the external potential considered in a typical LIF set-up, we can explicitly factor out the potential amplitude $V_0$ such that \mbox{$V_{\text{ext}} (x)= V_0 \cdot \tilde{V}_{\text{ext}} (x)$}.
The required potential amplitude~$V_0$ to enforce the relocation of particles from $L_s$ to $L_c$ (thus causing an increase from $\bar{\rho}$ to $\bar{\rho}_{\text{eff}}$) can then be explicitly calculated from Eq.~\eqref{Eq_abbreviated_EOS_form_averaged_stress_within_effective_density_ansatz} as
\begin{align}
 \beta V_0 &= \frac{Z(\bar{\rho}_{\text{eff}}) + \tilde{\Gamma}(\bar{\rho}_{\text{eff}}, L_c)}{\tilde{V}_{\text{ext}}\left(\frac{L_c}{2}\right)}.
 \label{Eq_potential_amplitude_V_0_to_enforce_compression_from_L_s_to_L_c}
\end{align}
For given $\makeCriticalEffectiveAverageDensitySymbol$, Eq.~\eqref{Eq_potential_amplitude_V_0_to_enforce_compression_from_L_s_to_L_c} yields the required potential amplitude $\beta V_0$ for the onset of LIF.
As stated earlier, we assume that spontaneous symmetry breaking occurs when the effective average density $\bar{\rho}_{\text{eff}}$ in the vicinity of the minima exceeds a critical value $\bar{\rho}_{\text{eff,c}}$.
A reasonable estimate of this critical value can be taken from the instability (with respect to freezing) of the corresponding bulk system (without $V_{\text{ext}}$).
For the hard disc bulk system, the freezing transition occurs at $\bar{\rho}_f \sigma^2 = 0.93$~\cite{Thorneywork2017}, and we take this value as the critical value for the onset of LIF, i.e., $\bar{\rho}_{\text{eff,c}}  \,\sigma^2=  0.93$.
The resulting prediction for the onset of LIF is shown as the red curve in Fig.~\ref{Fig_theoretical_prediction_for_phase_diagram}. We will discuss this curve in more detail in Sec.~\ref{Sec:Discussion_of_the_phase_diagram} in combination with the prediction for re-entrant melting (see below).


\subsection{Quantitative registration measure and re-entrant melting prediction \label{Sec:Quantitative_reentrant_melting_prediction}}

\begin{figure}
	\centering
	
	\includegraphics[width=0.49\textwidth]{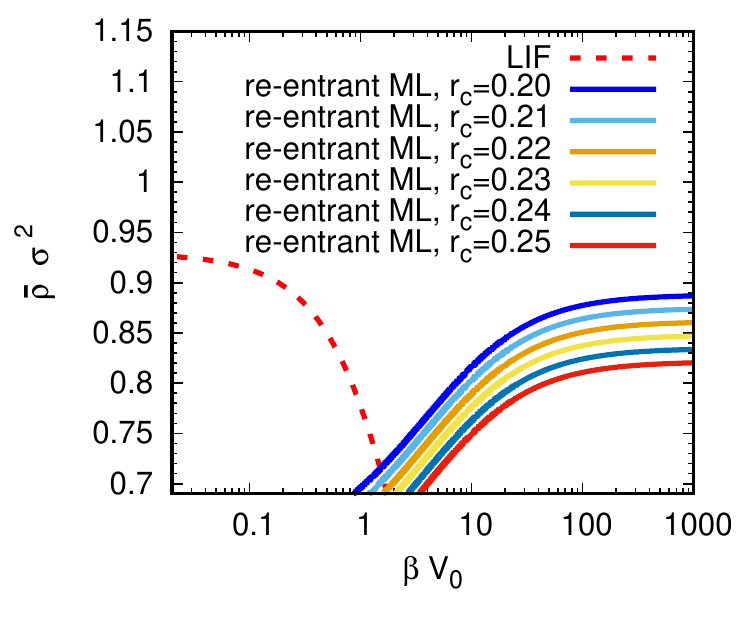}
	
	\caption{Re-entrant melting (ML) curves for different values of the critical value $r_c$ of the registration parameter $r$ [see Eq.~\eqref{Eq_Def_registration_measure}]. Also shown is our prediction for the onset of freezing (dashed line, see also Fig.~\ref{Fig_theoretical_prediction_for_phase_diagram}).
	Note that all melting curves are cut-off at the intersection point, since re-entrant melting should only occur after prior freezing.} 
	
	\label{Fig_vary_registration_threshold_for_interpolation_exponent_n_1}
\end{figure}

We now turn to the prediction for the re-entrant melting curve. Re-entrant melting is indeed a quite subtle effect, whose origin can be explained as follows~\cite{Wei1998}. In the locked floating solid phase, the fluctuations in $y$-direction (i.e., perpendicular to the potential barriers) are still quite large. These fluctuations are important for the mutual effective interaction between particles in adjacent minima. In particular, they contribute to the registration effect and are thus a crucial ingredient for the formation of the ordered phase. 
Upon further increase of $V_0$ (at given density $\bar{\rho}$), the potential barriers become larger and larger; leading to a decrease of particle correlations between adjacent minima and thus, to a reduction of the registration effect. In the most extreme case ($V_0\rightarrow \infty$), the 2D system is effectively reduced to 1D lines of particles which are known to have no positional order~\cite{Mermin1967, Mermin1968}.
Due to the role of fluctuations for re-entrant melting, it is not surprising that this phenomenon 
is not predicted by mean-field-like-theories (see comparison of mean-field-DFT~\cite{Chakrabarti1994} and MC simulation~\cite{Chakrabarti1995} studies). 
Within the present approach, the problem of describing re-entrant melting is even more severe because we are working with a parametrized density profile [see Eq.~\eqref{Eq_effective_density_ansatz}] where the density is described by only two parameters: the density inside the minima, $\makeEffectiveAverageDensitySymbol$, and the confining length $L_c$ measuring the actually accessible width of a minimum. 
As shown in our previous work where we numerically investigated LIF of ultra-soft spheres~\cite{Kraft2020a}, $L_c$ decreases with increasing $V_0$. This obviously implies that the difference $L_s-L_c$, with $L_s$ being the substrate periodicity, increases with $V_0$ as well (physically, $L_s-L_c$ corresponds to the excluded space). Moreover, our calculations in \cite{Kraft2020a} showed that when $L_s-L_c$ exceeds a certain fraction of the lattice constant $a$, there appears a gradual loss of correlations between adjacent lines; i.e., a reduction of registration. This observation motivates us to consider the gap parameter
\begin{equation}
\label{Eq_Def_registration_measure}
r= \frac{L_s- L_c}{a},
\end{equation}
as an indirect measure for the importance of perpendicular correlations. We further assume that there is a threshold value $r_c,$ beyond which the correlations are not sufficient any more to support the registration. Using Eq.~\eqref{Eq_Def_registration_measure}, this translates into a threshold value for the confining length $L_c$, that is, 
\begin{equation}
	\frac{L_c^{(r)}}{a} = \frac{L_s}{a} - r_c.
	\label{Eq_L_c_at_claimed_registration_threshold}
\end{equation}
Choosing $r_c$ appropriately, and inserting the resulting value for $L_c^{(r)}$ into Eq.~\eqref{Eq_potential_amplitude_V_0_to_enforce_compression_from_L_s_to_L_c} finally allows us to calculate the potential amplitude $V_0$, at which - in our framework - re-entrant melting sets in. Clearly, the remaining task is to choose the value for $r_c$. In Fig.~\ref{Fig_vary_registration_threshold_for_interpolation_exponent_n_1} we present results for the re-entrant melting curve for different values of $r_c$.
It is seen that variation of $r_c$ yields a monotonous shift of the entire curve, whereas the functional dependence on $V_0$ remains the same (this is also true for other choices of the exponent $n$ involved in the calculation of $\tilde{\Gamma}$, see Appendix~\ref{Sec:Impact_of_technical_parameters}).
In particular, any choice of $r_c$ leads to a saturation of the density values related to melting when $V_0$ becomes large. In view of this behaviour, we choose $r_c$ such that the limiting density $\bar{\rho}_\infty$ coincides with the corresponding value from MC simulations~\cite{Strepp2001,Chaudhuri2006}, $\bar{\rho}_\infty \sigma^2 = 0.9$. This calculation can be done numerically (by iteration). Alternatively, one can estimate $r_c$ directly from Eq.~\eqref{Eq_potential_amplitude_V_0_to_enforce_compression_from_L_s_to_L_c}. To this end we note that if $V_0\rightarrow\infty$, i.e., if the left side of Eq. ~\eqref{Eq_potential_amplitude_V_0_to_enforce_compression_from_L_s_to_L_c} diverges, the right side must diverge as well. This indeed happens when the density reaches its closed-packed limit, since then the compressibility factor diverges [see Eq.~\eqref{Eq_Compressibility_factor_bulk_system_hard_disc}]. Using Eq.~\eqref{Eq_redistribution_of_particles} with $\bar{\rho} = \bar{\rho}_\infty$ and $\makeEffectiveAverageDensitySymbol = \bar{\rho}_{\text{cp}}$, and $L_c = L_c^{(r)}$ given by Eq.~\eqref{Eq_L_c_at_claimed_registration_threshold}, a simple calculation yields $r_c \approx 0.19$. We henceforth take this value to calculate the re-entrant melting curve (see blue curve in the phase diagram  in Fig.~\ref{Fig_theoretical_prediction_for_phase_diagram}).


\section{Discussion of the phase diagram\label{Sec:Discussion_of_the_phase_diagram}}
\begin{figure}
	\centering
	
	\includegraphics[width=0.49\textwidth]{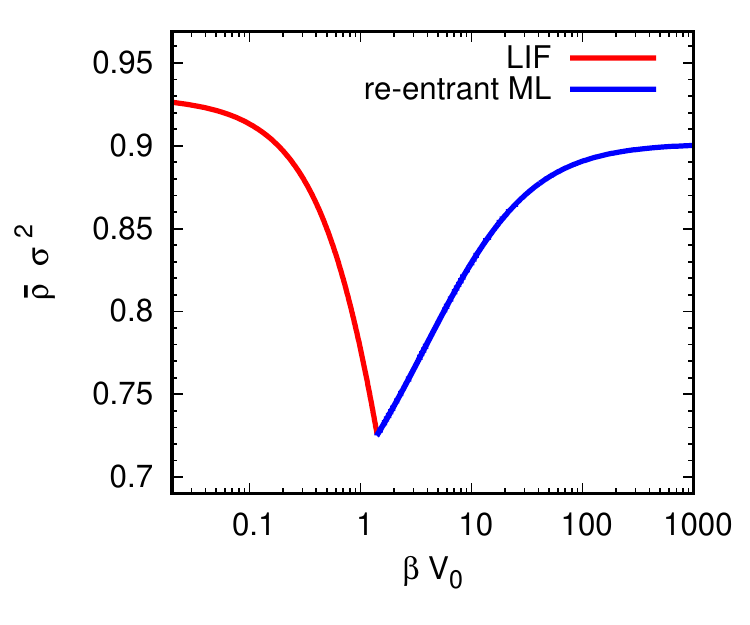}		
	\caption{Theoretical prediction for the onset of LIF and for re-entrant melting. 
		(The calculated curves have been cut-off at the intersection point, since re-entrant melting should only occur after prior freezing.)
	}
	\label{Fig_theoretical_prediction_for_phase_diagram}
\end{figure}

The full phase diagram consists of predictions for LIF and re-entrant melting as described in Secs.~\ref{Sec:Prediction_for_the_onset_of_LIF} and \ref{Sec:Quantitative_reentrant_melting_prediction}.
We now compare our calculated phase diagram, see Fig.~\ref{Fig_theoretical_prediction_for_phase_diagram}, with the phase diagram from the MC simulation study by Strepp \textit{et al.}~\cite{Strepp2001}~(see Fig.~\ref{Fig_phase_diagram_Strepp2001}).
Overall, we find good agreement, at least from a qualitative point of view.
Indeed, the calculated diagram accounts for important characteristics of the MC diagram:
Regarding LIF, the associated potential amplitude~$V_0$ goes to zero as the density $\bar{\rho}$ approaches the bulk freezing density. 
Upon decreasing $\bar{\rho}$, $V_0$ increases.
Further, the calculated re-entrant melting curve (see blue curve in Fig.~\ref{Fig_theoretical_prediction_for_phase_diagram}) displays a monotonous increase of the associated potential strengths with the density and reproduces the saturation observed in MC simulation in the limit $V_0 \to \infty$.  
The combination of the calculated LIF and re-entrant ML curves yields a minimum of the transition density at roughly the same potential amplitude, $\beta V_0 \approx 1-2$, as found in MC simulations~\cite{Strepp2001}. 
However, we remark that the actual value of this minimum density is underestimated in our approach.
Specifically, in the MC simulation~\cite{Strepp2001} the minimum was found at $\bar{\rho} \sigma^2 \approx 0.87$ ($\beta V_0 \approx 1-2$) while in our predicted phase diagram, the minimum occurs at $\bar{\rho} \sigma^2 = 0.73$ ($\beta V_0 = 1.4$).
Our calculated minimum transition density thus underestimates the value from MC simulations by roughly 16 \%.
This comparison shows that quantitative deviations in estimating $\bar{\rho}$ for given $V_0$ seem to be of the order of $16 \%$ or smaller (since both limits $V_0 \to 0$ and $V_0\to\infty$ agree exactly and phase boundary curves are monotonous).


\section{Conclusion\label{Sec:Conclusion}}

In the present paper, we employed a previously developed approach based on a pressure balance equation~\cite{Kraft2020a} to calculate the onset of freezing of a 2D system of hard discs subject to a 1D periodic substrate potential.
Thereby, one main goal was to utilize the approach such that it relies on bulk quantities, particularly, the bulk compressibility factor $Z$ and the bulk freezing density $\bar{\rho}_f$.
Specifically, to approximate the contribution to the pressure tensor due to inhomogeneity, we considered two limiting cases (i.e., $\bar{\rho} \to \bar{\rho}_f$ and $\bar{\rho}\to 0$) and connected them by a simple interpolation.
Finding further ways to exploit such limiting cases also for other types of interacting systems might be a potentially fruitful direction that is worth investigating. 
We here considered a hard disc system since hard-body interactions are often used as a reference for interacting many-body systems~\cite{HansenMcDonald}, and are thus particularly relevant. 
However, our hope is that the approach could also be applied to other systems, where the bulk behaviour is well understood.

Beyond freezing, we were able to make a prediction for the re-entrant melting that arises when increasing $V_0$ to values where registration between particles in neighbouring minima is hindered~\cite{Wei1998}, and the system dimensionality is effectively reduced to 1D channels. 
To this end, we defined a registration parameter $r$ [see Eq.~\eqref{Eq_Def_registration_measure}] and then claimed a threshold value~$r_c$ to arrive at a prediction for re-entrant melting.
In particular, we used the value $r_c = 0.19$, based on existing MC simulations~\cite{Strepp2001,Chaudhuri2006} in the limit of an infinite potential amplitude $V_0 \to \infty$. Knowing this limiting behaviour allows to predict re-entrant melting at \emph{finite} values of $V_0 < \infty$.

Combining the predictions for (laser-induced) freezing and re-entrant melting we obtained a phase diagram (see Fig.~\ref{Fig_theoretical_prediction_for_phase_diagram}), which shows unexpectedly good agreement with the phase diagram obtained by MC simulations~\cite{Strepp2001} (see Fig.~\ref{Fig_phase_diagram_Strepp2001}).
In view of the rather strong (yet reasonable) approximations, thereby enabling a calculation based on bulk quantities and known limiting behaviours, we would not expect exact quantitative agreement between the calculated phase diagram and the one obtained in Ref.~\onlinecite{Strepp2001}.
It is indeed surprising that the quantitative deviations in estimating $\bar{\rho}$ for given $V_0$ seem to be of the order of $16 \%$ or smaller.

Clearly, our present approach crucially foots on the exploitation of limiting cases where properties of the inhomogeneous system approach those of the homogeneous system. 
While this strategy seems fruitful in the present case, it is clearly important to perform more investigations for other types of interacting systems.
Also it seems worth investigating if the interpolation between those limiting cases is solely a technical aspect, or if more can be learned about the way the interpolation should be chosen.


\appendix

\section{Background of Eq.~\eqref{Eq_averaged_stress_within_effective_density_ansatz_variant3}\label{Sec:Background_of_our_framework}}

In this Appendix, we provide some background information from density functional theory which eventually leads to Eq.~\eqref{Eq_averaged_stress_within_effective_density_ansatz_variant3} (for a more detailed discussion, we refer to \cite{Kraft2020a}). 
The starting point is the exact balance equation of hydrostatics~\cite{Evans1979},
\begin{align}
	\nabla \cdot \bs{\sigma}(\bs{r}) = \rho(\bs{r}) \nabla V_{\text{ext}}(\bs{r}),
\label{Eq_Cauchy_momentum_equation_stationary}
\end{align}
where $\bs{\sigma}$ denotes the (second-order) stress tensor, which is the negative of the usual pressure tensor~\cite{Henderson1992}.
Physically, Eq.~\eqref{Eq_Cauchy_momentum_equation_stationary} states that the stress inside the system balances the force stemming from the external potential $V_\text{ext}$.
More formally, Eq.~\eqref{Eq_Cauchy_momentum_equation_stationary} shows that the divergence of the stress tensor $\bs{\sigma}$ is a functional of the density profile~$\rho(\bs{r})$ (see Ref.~\cite{Kraft2020a} for more details).
We note that in the limit of a constant or vanishing external potential [$V_\text{ext}(x) \to 0$ or $\nabla V_\text{ext} \to 0$], both sides of Eq.~\eqref{Eq_Cauchy_momentum_equation_stationary} become zero (no net external forces and zero divergence of the stress tensor). 
Given the true density profile $\rho(\bs{r})$ and the true correlations in the system, Eq.~\eqref{Eq_Cauchy_momentum_equation_stationary} is exact.

However, the true density profile  $\rho(\bs{r})$ is typically not known.
The approach that we follow (see \cite{Kraft2020a}) is to consider a spatially integrated version of Eq.~\eqref{Eq_Cauchy_momentum_equation_stationary}.
In particular, we integrate Eq.~\eqref{Eq_Cauchy_momentum_equation_stationary} over an area $\mathcal{A}$ in the $x$-$y$ plane, and divide the resulting integrals by that area, yielding 
\begin{align}
\label{Eq_averaged_stress}
	\frac{1}{\mathcal{A}} \int d\mathcal{A}\, \text{sign}(x)\, &\bs{e}_x \, \nabla \cdot\bs{\sigma} \notag\\
	&= \\ 
	\frac{1}{\mathcal{A}} \int d\mathcal{A}\, \text{sign}(x)\, &\bs{e}_x \rho(\bs{r}) \nabla V_{\text{ext}}(x) ,\notag
\end{align}
where $\bs{e}_x$ denotes the unit vector in $x$-direction and $\text{sign}(\cdots)$ denotes the sign function.
The region $\mathcal{A} = [-\frac{L_x}{2}, \frac{L_x}{2}] \times [-\frac{L_y}{2}, \frac{L_y}{2}]$ is chosen to be centered around the minimum of the substrate potential $V_{\text{ext}}(x)$, say $x=0$ (for notational convenience).
We note that the quantities involved in Eq.~\eqref{Eq_Cauchy_momentum_equation_stationary} are anti-symmetric with respect to the location of the minimum, such that a direct average would result to zero.
We thus multiply both sides of Eq.~\eqref{Eq_Cauchy_momentum_equation_stationary} by $\text{sign}(x)$.
We further multiply with $\bs{e}_x$, since we are interested in the $x$-component of the force.
Equation~\eqref{Eq_averaged_stress} is still exact, providing a starting point for approximations \cite{Kraft2020a}. Here we use, in particular, a simple ansatz for the density (and pressure) profile.

Due to this strong approximation several variants are possible to evaluate Eq.~\eqref{Eq_averaged_stress}~\cite{Kraft2020a}.
We here work with a variant which worked best for ultrasoft particles~\cite{Kraft2020a} and that we termed variant B.II.
In addition to the rectangular ansatz for the density profile [see Eq.~\eqref{Eq_effective_density_ansatz}], we also made a rectangular ansatz for the pressure profile according to	
	\begin{alignat}{2}
	p(x)  	&=p(\bar{\rho}_{\text{eff}} )\, \text{rect}\left( \frac{x}{L_c} \right),
	\label{Eq_effective_pressure_profile_ansatz}
	\end{alignat}
where $p(\bar{\rho}_{\text{eff}} )$ is the pressure of a \textit{bulk} system of constant density $\makeEffectiveAverageDensitySymbol$. 
Evaluating Eq.~\eqref{Eq_averaged_stress} in variant B.II yields Eq.~\eqref{Eq_averaged_stress_within_effective_density_ansatz_variant3}.

\section{Impact of technical parameters\label{Sec:Impact_of_technical_parameters}}

\begin{figure}
	\centering
	
	\includegraphics[width=0.49\textwidth]{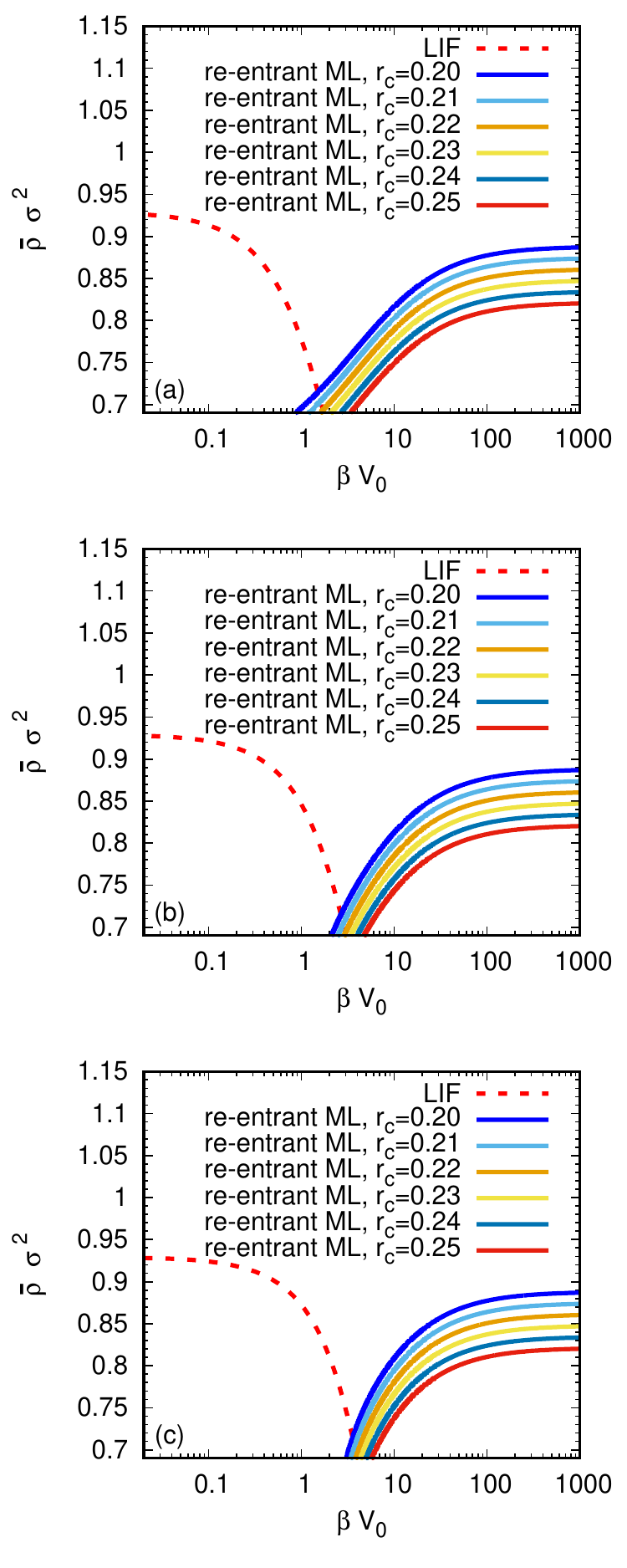}

	\caption{
		Predictions of the phase diagram for different exponents~$n$ of the interpolation for $\tilde{\Gamma}$ [see Eq.~\eqref{Eq_Interpolation_for_GammaTilde}] and for different values of the critical value $r_c$ of the registration parameter $r$ [see Eq.~\eqref{Eq_Def_registration_measure}].
		Parts (a)-(c) are for the exponents $n=1,2,3$, where the results for $n=1$ are identical to those in Fig.~\ref{Fig_vary_registration_threshold_for_interpolation_exponent_n_1}.
		(Note that all curves are cut-off at the intersection point, since re-entrant melting should only occur after prior freezing.)} 
	
	\label{Fig_vary_interpolation_exponent_and_registration_threshold}
\end{figure}

In this Appendix we discuss different choices for some technical parameters required for our calculation of the LIF prediction and re-entrant melting prediction (see sections \ref{Sec:Prediction_for_the_onset_of_LIF} and \ref{Sec:Quantitative_reentrant_melting_prediction}).
This concerns, in particular, the exponent $n$ appearing in the interpolation $\left({\bar{\rho}}/{\bar{\rho}_f}\right)^n$ in Eq.~\eqref{Eq_Interpolation_for_GammaTilde}, and the parameter $r_c$ appearing in Eq.~\eqref{Eq_L_c_at_claimed_registration_threshold}.
Results for different choices of these parameters are shown in Fig.~\ref{Fig_vary_interpolation_exponent_and_registration_threshold}.
The three parts correspond to the exponents $n$ = 1, 2, and 3 (we here included the data for $n=1$ already shown in Fig.~\ref{Fig_vary_registration_threshold_for_interpolation_exponent_n_1} to enable a better comparison.)
The following observations can be made.
Regardless of $r_c$, the minimum (intersection point of curves for LIF prediction and re-entrant melting prediction) is shifted to larger values of $\beta V_0$ for larger $n$.
Further, we see that the influence of the exponent $n$ becomes stronger for increasing distance of $\bar{\rho}$ values from the bulk freezing density $\bar{\rho}_f$. 
This is somewhat expected because our goal was to make a prediction based on bulk information. Indeed, for densities close to the bulk freezing density, the value of the exponent $n$ seems to have a negligible influence.
Altogether, the general topology of the phase diagram seems quite insensitive to the exponent.

In Fig.~\ref{Fig_vary_interpolation_exponent_and_registration_threshold}, we also show re-entrant melting predictions for different values of the critical value $r_c$ of the registration parameter $r$ [see Eq.~\eqref{Eq_Def_registration_measure}].
We observe that increasing $r_c$ shifts the re-entrant melting curve downwards.

\bibliography{myReferences_paper2}{}

\end{document}